\documentclass[10pt,twocolumn]{article}
\usepackage[T1]{fontenc}
\usepackage{color,times,graphicx}
\usepackage{hyperref}
\usepackage{algorithm,amsmath,amsfonts}
\usepackage[noend]{algorithmic}
\usepackage{xspace}

\usepackage[medium,compact]{titlesec}

\long\def\eat#1{}

\newcommand{\Sys}{\textsc{VPM}\xspace}
\newcommand{\HOP}{HOP\xspace}
\newcommand{\HOPs}{HOPs\xspace}

\newcommand{\dlp}{\mathcal{P}}
\newcommand{\packet}{\mathit{p}}
\newcommand{\agg}{\mathit{\alpha}}
\newcommand{\aggb}{\mathit{\beta}}
\newcommand{\stream}{\mathcal{S}}
\newcommand{\seq}{\mathcal{\hat{S}}}
\newcommand{\aggset}{\mathcal{A}}

\newcommand{\aggsetj}{\mathcal{J}}
\newcommand{\receipt}{\mathcal{R}}

\newcommand{\pathID}{\mathit{PathID}}
\newcommand{\headerSpec}{\mathit{HeaderSpec}}
\newcommand{\previousHOPID}{\mathit{PreviousHOP}}
\newcommand{\nextHOPID}{\mathit{NextHOP}}
\newcommand{\aggID}{\mathit{AggID}}
\newcommand{\firstPacket}{\mathit{FirstPacketID}}
\newcommand{\lastPacket}{\mathit{LastPacketID}}
\newcommand{\pktCount}{\mathit{PktCnt}}

\newcommand{\packetID}{\mathit{PktID}}
\newcommand{\timestamp}{\mathit{Time}}
\newcommand{\checkValue}{\mathit{MaxDiff}}
\newcommand{\aggTrans}{\mathit{AggTrans}}

\newcommand{\dsamp}{\mathit{DelaySample}} 
\newcommand{\marker}{\mathit{\mu}} 
 
\newcommand{\samplingthresh}{\mathit{\sigma}} 
\newcommand{\samplingFunction}{\mathit{SampleFcn}} 
\newcommand{\samples}{\mathit{Samples}}

\newcommand{\join}{\mathit{Join}} 
 
\newcommand{\descthresh}{\mathit{\delta}} 
\newcommand{\descriptor}{\mathit{Digest}} 
 
\newcommand{\maxJitter}{\mathit{J}}

\renewenvironment{thebibliography}[1]{%
    \begin{oldthebibliography}{#1}%
    \setlength{\parskip}{0ex}%
    \setlength{\itemsep}{0ex}%
}%
{%
    \end{oldthebibliography}%
}

\def\compactify{\itemsep=0pt plus3pt \topsep=3pt plus3pt \partopsep=0pt
\parsep=0pt \leftmargin=\parindent \labelwidth=\leftmargin}
\let\latexusecounter=\usecounter
\def\CompactItemize{%
  \ifnum \@itemdepth >\thr@@\@toodeep\else
    \advance\@itemdepth\@ne
    \edef\@itemitem{labelitem\romannumeral\the\@itemdepth}%
    \expandafter
    \list
      \csname\@itemitem\endcsname
      {\compactify\def\makelabel##1{\hss\llap{##1}}}%
  \fi}

\newenvironment{CompactEnumerate}
  {\def\usecounter{\compactify\latexusecounter}
   \begin{enumerate}}
  {\end{enumerate}\let\usecounter=\latexusecounter}

\begin{document}
\title{
\vspace{-1.2em}
\bf\Large Verifiable Network-Performance Measurements
\vspace{-0.5em}
}
\author{ 
%Paper \#134, 14 pages
\vspace{-0.5em}
\begin{tabular}{c c c}
Katerina Argyraki & Petros Maniatis & Ankit Singla \\
EPFL, Switzerland & Intel Research Berkeley & EPFL, Switzerland \\
\end{tabular}
}
\date{}
\maketitle

\begin{abstract}
In the current Internet, there is no clean way for affected
parties to react to poor forwarding performance: when a domain violates
its Service Level Agreement (SLA) with a contractual partner, 
the partner must resort to ad-hoc probing-based monitoring to determine the
existence and extent of the violation.  
Instead, we propose a new, systematic approach to the problem of forwarding-performance verification.  
Our mechanism relies on voluntary reporting, allowing each domain to disclose its loss and delay performance to its customers and peers.
Most importantly, it enables \emph{verifiable} performance measurements, i.e.,
domains cannot abuse it to significantly exaggerate their performance.
Finally, our mechanism is \emph{tunable}, allowing each participating domain to determine how many resources to
devote to it independently (i.e., without any inter-domain coordination),
exposing a controllable trade-off between performance-verification
quality and resource consumption.
Our mechanism comes at the cost of deploying modest functionality at the
participating domains' border routers; we show that it requires 
reasonable resources, well within modern network capabilities.
\end{abstract}

\section{Introduction}
\label{sec:intro}

The lack of a systematic method for estimating the performance of Internet service providers (ISPs) is a well known problem:
when an ISP does not perform as expected, there is no clean way for the affected parties to detect the problem so they can debug it, 
ask for compensation if a Service-Level Agreement (SLA) has been violated, or simply learn from it (e.g., re-assess 
a peering agreement with an under-performing neighbor).  
This lack of information makes network debugging difficult and slow, even leading ISPs to deny their failures to their customers and peers, pointing fingers at one another.
One could attribute this situation to the best-effort nature of the Internet which, by definition, provides no a-priori guarantees. 
Yet that is no reason not to expect useful, after-the-fact information about ISP performance---actually, it makes perfect sense to expect such information 
in a best-effort environment like the Internet, where communication quality often relies on quick failure detection and on choosing the right providers and peers.  

Since ISPs offer no explicit interface for their customers and peers to verify their performance, the latter can only resort to probing tools
like traceroute or other active measurements.
Moreover, researchers have recently started to combine probing from multiple vantage points (e.g., PlanetLab nodes) to gain information 
about ISP performance that would not be accessible through simple probing~\cite{Katz08, Katz10}.
This information is typically extracted from channels with a different purpose (e.g., ICMP traffic),
because probing mechanisms are designed under the assumption that ISPs would never freely provide honest information about their
performance. 

But what if ISPs \emph{were} willing to export an explicit interface through which their performance can be queried?
In this work, we ask the question, how should we design such an interface such that it provides accurate and verifiable information, 
while it can be implemented using a reasonable, tunable amount of resources?
On the one hand, we find this to be an interesting thought experiment.
On the other hand, we identify two strong, albeit perhaps unintuitive, reasons why 
an ISP may willingly expose its performance problems to the outside world.

First, ISPs often need to exchange performance information anyway with their customers and peers, in order to handle customer complaints.
When a customer calls her ISP to complain that she cannot reach a certain destination, the ISP needs to know whether the problem
lies in its own local network, the customer's network, the network of the peer that is handling traffic to that destination, or the destination's network---because each
of these cases warrants a different response.
Today, this information is acquired by ISP operators in a reactive, ad-hoc manner, which means that it takes time to resolve each complaint,
potentially leaving customers dissatisfied.
It makes sense that an ISP would prefer to collaborate with its customers and peers and willingly exchange troubleshooting reports with them,
provided that it can trust these reports to be accurate and honest.

Second, it makes sense that an ISP would prefer to report its own
performance rather than have its performance evaluated by untrusted
entities, through potentially inaccurate mechanisms.  Probing or other
edge-based ``black-box'' mechanisms typically run on coalitions of
end-systems like PlanetLab; the ISP has no reason to trust these,
and they can provide no guarantee for the accuracy of their measurements.       
If an ISP's performance is to be talked about anyway, an accurate, trusted
self-reporting mechanism may be preferable to the ISP, because, at
least, it provides the ISP with control over the quality and quantity of
the information that is revealed about its business.

Self-reporting is not necessarily better or worse than edge-based probing; each approach has different pros and cons.
On the one hand, while probing is effective for localizing persistent outages or high-rate drop patterns, 
it provides no reliable indicator of the fate of non-probe traffic: probes can be treated differently, either by design (e.g.,
ICMP packet responses are generated off the fast path of routers), or by ``strategic thinking'' (treating probe packets preferentially 
to improve externally perceived performance). 
On the other hand, whereas probing is simple and requires no changes in ISPs, a self-reporting mechanism by
necessity requires some extra complexity in the control- and data-plane mechanisms of the Internet's forwarding fabric. 

In the rest of the paper, we describe a self-reporting mechanism for verifiable network-performance measurements (or \Sys, for brevity).
According to \Sys, each ISP's loss and delay performance is cooperatively estimated by the ISP itself and
the other network domains (customers and peers) that carry its traffic.  
Its key features are:
(1) It enables accurate estimation of ISP performance, without revealing any information about
the internal structure or routing policies of ISPs beyond what is already publicly available through BGP routing tables.
(2) ISPs cannot abuse it to significantly exaggerate their performance.
(3) It allows each ISP to choose its own cost/quality trade-off independently from others, yet in a way that does not
compromise the verifiability of the derived measurements.
These features come at the cost of deploying new functionality at the participating domains' border routers,
but we show that the corresponding memory, processing, and bandwidth requirements are well within the
capabilities of modern networks.

We start, in Section~\ref{sec:setup}, with a high-level description of our approach, 
followed by a more precise problem statement and our assumptions.
Section~\ref{sec:new} explains why existing protocols or straightforward combinations of existing techniques fail to provide an appropriate solution.
Section~\ref{sec:reporting} describes what kind of information \Sys collects and disseminates among participating domains. 
Sections~\ref{sec:sampling} and~\ref{sec:aggregation} describe how \Sys provides independent tunability of resource expenditure at different domains 
while still achieving high quality of information. 
Section~\ref{sec:eval} evaluates \Sys experimentally and through back-of-the-envelope calculations, in terms of its overhead and information quality provided.
Section~\ref{sec:related} discusses partial deployment and related work,
and Section~\ref{sec:conclusions} concludes.

\section{Setup}
\label{sec:setup}

In this section, we first describe our approach at a high level (\S\ref{sec:setup:approach}),
then provide a more concrete problem statement (\S\ref{sec:setup:statement}) and state our assumptions (\S\ref{sec:setup:assumptions}).

\begin{figure}[t]
\centering
\includegraphics{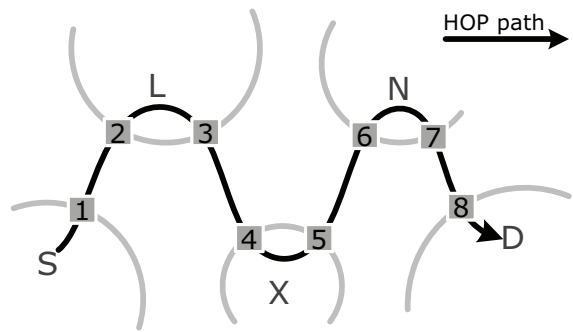}
\caption{\small Circles represent administrative domains. The numbered boxes represent \HOPs.
The black arrow represents a \HOP path. Our main example scenario throughout the paper:
domain $S$ sends to domain $D$ a packet set $\stream = \{p_1, p_2, ...\}$ via \HOPs 1 to 8.}
\label{fig:sla}
\vspace{-0.3cm}
\end{figure}

We will use the following terminology.
A ``domain'' is a contiguous network that falls under one administrative entity;
in the current Internet, a domain would refer to an edge network or a single Autonomous System (AS).
Each domain has \emph{hand-off points} (or \emph{\HOP}s) along its perimeter; 
these are ingress/egress points, where traffic enters/exits the domain's jurisdiction (see Figure~\ref{fig:sla} for examples).
Each \HOP is connected to a neighboring domain's \HOP through an inter-domain link; 
such a link is considered \emph{faulty} when it introduces loss or delay beyond a known specification. 
We are in particular interested in packets traversing the same \emph{\HOP path}, i.e., the same sequence of \HOPs;
we name such paths according to their source and destination routing prefixes (that is, origin prefixes as advertised in BGP).

\subsection{Approach}
\label{sec:setup:approach}

In \Sys, each domain monitors traffic at its \HOPs and produces \emph{receipts} for the traffic that enters and exits its network.
For privacy reasons, a receipt is made available only to the domains that observed the corresponding traffic.
For instance, if any of the domains in Figure~\ref{fig:sla} produces a receipt for a set of packets $\{p_1, p_2, ...\}$ that crossed domains
$S$, $L$, $X$, $N$, and $D$, the receipt is made available only to these particular 5 domains.
To ensure this, each \HOP classifies observed traffic per \HOP path and produces a common receipt only for packets that followed the same \HOP path.
This implies that when a \HOP observes two packets $p_1$ and $p_2$, the \HOP knows (in practice, can guess with a high probability) whether the two packets
belong to the same \HOP path (see Assumption \#1 below).

Each domain $X$ collects receipts from its neighbors with the purpose of estimating each neighbor's loss and delay performance with respect to its traffic.
Moreover, domain $X$ collects receipts from the other domains that observed its traffic with the purpose of verifying the correctness of its neighbors' receipts.
The idea is that if a neighbor provides incorrect receipts to exaggerate
its own performance (e.g., claim that it delivered traffic that it actually dropped),
these ``dishonest'' receipts will be inconsistent with the receipts of the other domains on the path.

We do not worry, in this paper, about how or when receipts are disseminated (see Assumption \#2 below).
A domain could request receipts periodically (e.g., once an hour or once a day) or arrange to receive them in real time, as they are generated.
Collecting receipts from \emph{all} other domains that handle a domain's traffic may sound like overkill at first---and it would be, if receipts were produced per packet or per flow.
However, in \Sys, receipts are produced at coarser granularity, such
that each domain incurs, due to receipts, less than $0.1$\% overhead
over the traffic it observes (\S\ref{sec:eval}).

Instead, we focus on the content of the receipts.
We ask the question, if domains were willing to provide receipts on the traffic they receive and deliver, what should these receipts consist of, such
that (i) they can be generated using a reasonable, tunable amount of resources and (ii) neighbors can use them to estimate and verify each other's performance?

\paragraph*{Threat Model} 
We assume the existence of both \emph{honest} domains that construct their receipts exactly as our protocol specifies
and \emph{lying} domains that construct their receipts using incomplete or fabricated information.
Our threat model allows lying domains to \emph{collude} with others towards a common nefarious goal.  
Nevertheless, a lying domain can observe only network traffic that appears locally (because it originates at, 
terminates at, or transits that domain), or that has been observed by its colluding domains.

We do not consider, in this paper, the scenario where domains modify observed traffic.
This is not because this scenario is not plausible or not interesting, but because it is, to the best of our
knowledge, further from current ISP practices (than introducing loss or unpredictable delay and denying performance problems).
Moreover, as we will see, dealing with loss and delay without considering traffic modification 
is already a challenging enough problem to warrant separate treatment.

\subsection{Problem Statement}
\label{sec:setup:statement}

Consider a path $\dlp$, like the one pictured in Figure~\ref{fig:sla}.
Suppose that each \HOP in $\dlp$ can disseminate a certain amount of information to all other \HOPs in $\dlp$.
The question is, what should this information be, such that the following conditions are met:

\paragraph*{1. Computability}
As long as domain $X$ in path $\dlp$ produces honest information, $X$'s neighbors in $\dlp$
can use that information to compute the loss and delay introduced by $X$ in the traffic flowing along $\dlp$.

Regarding delay, we are interested in delay quantiles, e.g., domain $L$ should be able to determine that domain 
$X$ introduced delay below $5$msec to $90$\% of the traffic with a certain (high) probability $\pi$.
We are interested in quantiles, \emph{not} delay averages, because a domain may exhibit low average delay
at the time scale of seconds or minutes, yet introduce ``spikes'' of high delay that can impact 
the performance of TCP or real-time applications significantly~\cite{Markopoulou06}.

\paragraph*{2. Verifiability}
If domain $X$ in path $\dlp$ produces dishonest information, its neighbors in $\dlp$ can detect and discard that information.

\paragraph*{3. Tunability}
The amount of resources consumed in collecting and disseminating information is locally \emph{tunable} by each \HOP,
such that the accuracy of the statistics computed from this information degrades \emph{gracefully}
with the amount of resources spent to collect and disseminate it.

\subsection{Assumptions}
\label{sec:setup:assumptions}

In addition to our threat model, we make the following assumptions:

(1) Our strongest assumption is that the \HOP path over which traffic
between the same source and destination origin prefix is routed changes
only slowly (i.e., on the order of hours, rather than seconds).  This is
largely the case today for domain-level paths over short time
scales.  Note that this does \emph{not} restrict how a domain
load-balances traffic internally. Each domain is free to split traffic
through multiple internal paths in any way it wants, as long as it
forwards all traffic with the same source/destination prefixes via the
same egress link.

(2) We assume that there exists a way for a domain in path $\dlp$
to disseminate receipts to all other domains in $\dlp$, such that the
authenticity and integrity of each received receipt is guaranteed.  
One way of realizing this assumption would be for each domain to make its receipts available
at an administrative web-site and accessible over HTTPS.
It is possible to design more efficient dissemination mechanisms, but that is outside the scope of this paper.

(3) Finally, we assume that each domain has some network equipment (routers or other
middleboxes) that can perform at wire speed simple per-packet
operations. Those include packet timestamp generation, arithmetic
calculations or digest computations on packet headers and a small
portion of packet payload, and modification of local state in a buffer.
This assumption is well
justified by current trends in production routers, as well as the
increasing focus of academia and industry on programmable routers and
switches~\cite{OpenFlow, RouteBricks}.

\section{Why a New Protocol}
\label{sec:new}

There already exist many good techniques for measuring network performance~\cite{Duffield01,Sommers07,Goldberg08,Kompella09}.
So, instead of describing \Sys from scratch, we first build, in this section, ``obvious'' solutions by combining or 
extending existing techniques, and describe why each of these solutions fails to meet the three conditions of our problem statement.
We close with an overview of \Sys and how it relates to the existing techniques.

\subsection{Strawman}
\label{sec:new:strawman}

As a first-cut, strawman solution,
we consider the following modest extension to the Packet Obituaries protocol~\cite{Maniatis04}:
Each \HOP produces a receipt for every single packet it observes. 
A receipt consists of a \emph{digest} for the corresponding packet and the \emph{timestamp} for when the packet was observed.
Each receipt is made available to \emph{all} the domains that observed the packet.

\paragraph*{Computability}
The strawman easily meets this condition, as a receipt collector in possession of all the (honest) 
receipts generated by a domain $X$ can determine
whether each packet that entered $X$ was dropped within $X$ and, 
if not, by how much it was delayed within $X$.
By combining such information for multiple packets, the receipt collector can easily compute aggregate loss statistics
and delay quantiles for $X$.

\paragraph*{Verifiability}
The strawman also meets this condition:

To hide a loss or delay incident, a domain has to falsely put the blame for the incident on one of its neighbors,
which results in inconsistent claims between the two domains.
For instance, suppose domain $X$ receives packet $p$ from domain $L$ but drops it before delivering it no domain $N$.
If $X$ is dishonest and wants to hide the fact that it dropped $p$,
it can put the blame on $N$, i.e., falsely claim having delivered $p$ to $N$. 
This claim will be inconsistent with $N$'s claim of not having received $p$.

Such an inconsistency can be due either to a lie or to a faulty inter-domain link.  
If a receipt collector receives inconsistent claims from two neighbors, it discards the corresponding
receipts (from both neighbors) and notifies both of them of the inconsistency.  
The two involved neighbors can then debug their inter-domain link; if it is functioning correctly, 
then the inconsistency was due to a lie, and the lying domain is exposed to the neighbor it implicated.
For instance, if $X$ falsely reports having delivered packet $p$ to $N$,
but $N$ correctly reports not having received $p$, the rest of the world cannot determine whether $X$ or $N$ is lying,
but $N$ does know that $X$ is the liar.

A domain can always support a lying neighbor's claims, but then it either
has to take itself the blame for the liar's loss/delay or falsely accuse 
another domain down the path.
For instance, if $X$ falsely claims having delivered $p$ to $N$,
$N$ has the option of covering $X$'s lie (by claiming that it indeed received $p$),
but then it has to claim either that it lost $p$ itself, or that it delivered $p$ to
$D$---in which case $N$ is exposed to $D$ as a liar.

It is important to note that the strawman meets the verifiability constraint, only because each receipt collector
collects receipts from \emph{all} \HOPs on the path, and computes the performance of \emph{all} domains.
If, instead, each receipt collector collected receipts only from a segment of the path, 
then there would be no incentive for domains to be honest about their neighbors' performance.
For instance, suppose domain $L$ wants to compute domain $X$'s performance but collects receipts only from \HOPs 3, 4, and 5.
Suppose domain $X$ drops packet $p$ and falsely claims having delivered $p$ to $N$.
In this case, $N$ can safely cover $X$'s lie, i.e., claim having received $p$.
Since domain $L$ does not collect receipts beyond \HOP 5, it has no way of computing $N$'s performance
and verifying it against $D$'s receipts. Hence, $N$ can collude with $X$ and cover its lie without any harm
to its own reputation.

\paragraph*{Tunability}
This is where the strawman fails.
The cost of maintaining and propagating per-packet receipts, though not intractable, 
can be expensive in buffering space, processing, and reporting bandwidth. Different
domains may have different resources they are willing to devote
to a self-reporting endeavor, and keeping per-packet state leaves no room for tuning.

\subsection{Trajectory Sampling ++}

Since the main problem with the strawman is the non-tunable cost of collecting and exchanging per-packet state,
the first solution that comes to mind is to sample, i.e., collect information not on all packets, 
but on a representative subset, and use it to infer statistics for the rest.
Hence, we next consider a combination of the strawman and Trajectory Sampling~\cite{Duffield01}
(we call it ``Trajectory Sampling ++'').

Each \HOP applies a uniform hash function to a small, fixed portion of each observed packet.
If the outcome exceeds a pre-configured threshold, then the packet is sampled and the \HOP produces a receipt for it.
Each pair of \HOPs from the same domain use the same hash function and sampling threshold, hence sample the same packets.
Each receipt is made available to all the domains that observed the corresponding packet.

\paragraph*{Computability}
This condition is met, both for loss and delay statistics.
First, a receipt collector in possession of all the (honest) receipts produced 
by a domain $X$ can count how many of the sampled packets were lost within $X$;
from that, it can estimate how many packets were lost within $X$ overall, as shown in~\cite{Sommers07}.
Similarly, the receipt collector can compute the delay incurred by each sampled packet within $X$,
then estimate delay quantiles for the overall traffic~\cite{Sommers07}.

\paragraph*{Verifiability}
This is where Trajectory Sampling ++ fails,
and we will argue that this failure is inherent to any sampling-based solution.

The obvious problem with sampling is that a domain can lie about its performance by biasing the sampling process.
Since a domain's performance is estimated based on how it treats the sampled packets,
if domain $X$ treats the sampled packets preferentially (i.e., assigns them to high-priority queues),
then $X$'s estimated performance will be higher than its actual performance. 

On a first thought, such cheating seems easy to detect, as long as not all \HOPs sample the same packets.
We illustrate with an example.
Suppose \HOPs 4 and 5 from Figure~\ref{fig:sla} sample one set of packets, $s_1$,
whereas \HOPs 3 and 6 sample a \emph{different} set of packets, $s_2$.
Suppose domain $L$ wants to estimate domain $X$'s performance and collects receipts from all \HOPs.
First, $L$ uses the receipts from \HOPs 4 and 5 to estimate the loss and delay incurred between these two \HOPs.
Similarly, $L$ uses the receipts from \HOPs 3 and 6 to estimate the loss and delay incurred between them.
If the two sets of statistics do not match (e.g., the estimated loss between \HOPs 4 and 5 is significantly lower than
the estimated loss between \HOPs 3 and 6), then: either one or both of the involved inter-domain links are malfunctioning, 
or domain $X$ is biasing its samples to exaggerate its performance,
or domain $N$ is biasing \emph{its} samples to misrepresent $X$'s performance.
Hence, one could argue, as long as not all \HOPs sample the same packets (hence, not all \HOPs have a reason to
bias the same traffic), we can get similar incentives with the strawman, 
i.e., lies lead to inconsistencies, and liars are exposed to their neighbors.

The main problem with this argument is that it assumes that domain $X$ 
(i.e., \HOPs 4 and 5) treats the packets from set $s_1$ preferentially, but the packets from $s_2$ normally 
(like the rest of the traffic);
yet there is a clear incentive here for domains $X$ and $N$ to collude and treat \emph{both} sets of sampled packets preferentially,
such that they make consistent claims, 
\emph{and} the statistics computed from their receipts overestimate the performance of both of them.
There are also other problems, less fundamental, but potentially significant in practice:
This approach requires \HOPs from different domains (in our example, \HOPs 3 and 6)
to agree to sample the same packets. Moreover, an ``inconsistency'' is now a difference in a probabilistic
estimate---not a concrete disagreement about a particular packet as in the strawman.

To conclude, when each domain's performance is estimated based on how it treats sampled packets,
then a sequence of interconnected domains have an incentive to collude and bias the samples taken by all of them.
In contrast, when domains provide receipts for every single packet, there is no incentive for such misbehavior,
because colluding with a neighbor to cover the neighbor's failures necessarily means taking the blame yourself.

An explanation of why the ``Secure Sampling'' technique from~\cite{Goldberg08} does not address this problem can
be found in Section~\ref{sec:related}.

\subsection{Difference Aggregator ++}

An alternative way of introducing tunability in the strawman is to aggregate,
i.e., collect information not for individual packets, but for groups of packets.
The benefit of aggregation versus sampling is that each domain produces information that depends
on \emph{all} the packets it observes, hence there is no straightforward way to cheat by treating certain packets preferentially.
Hence, we next consider the following combination of the strawman and
Lossy Difference Aggregator\footnote{We could have equally considered a combination of the strawman and the 
``Secure Sketch'' technique from~\cite{Goldberg08}. The conclusion would have been the same. For a comparison with that work, 
see Section~\ref{sec:related}.}~\cite{Kompella09} (we call it ``Difference Aggregator ++'').

Each \HOP breaks the sequence of observed packets from a given path into packet aggregates, 
where a ``packet aggregate'' is a set of consecutively observed packets.
For example, if a \HOP observes packet sequence $\langle \packet_1, \packet_2, \packet_3, \packet_4, \packet_5 \rangle$\footnote{In 
reality, a \HOP would observe infinite packet sequences. In our examples, we use finite sequences for simplicity.}
from path $\dlp$, it may break that into two aggregates $\{\packet_1, \packet_2, \packet_3\}$ and $\{\packet_4, \packet_5\}$.
For each aggregate, the \HOP computes a packet count and an average timestamp, and stores them in a receipt, 
together with an identifier for the aggregate.
Each receipt is made available to all domains that observed the corresponding aggregate.

Moreover, each pair of \HOPs from the same domain try to break the observed traffic into the same set of aggregates.
A classic approach is to use common ``cutting points'':
Each \HOP applies a uniform hash function to a small, fixed portion of each observed packet.
If the outcome is larger than a pre-configured threshold, 
then the packet is considered a ``cutting point'' and starts a new packet aggregate.
If two \HOPs use the same hash function and cutting threshold, and there is no packet re-ordering between them,
then the two \HOPs end up breaking the observed traffic into the same set of packet aggregates.

\paragraph*{Computability}
Difference Aggregator ++ fails to meet the computability condition in two ways.
First, it cannot provide meaningful statistics in the face of packet reordering.
Second, even if there is no packet reordering, it cannot provide sufficient information for estimating delay 
quantiles---only for computing loss and estimating average delay.

Let's assume, temporarily, that there is no packet reordering within domain $X$.
In this case, a receipt collector in possession of the (honest) receipts produced by $X$ can compute the loss
incurred by each packet aggregate $\alpha$ within $X$, by comparing the packet counts collected for $\alpha$ at \HOPs 4 and 5.
By combining such information for multiple aggregates, one can precisely compute the loss incurred by the overall traffic within $X$.
Less obviously, by taking into account only the aggregates that did not incur any packet loss,
one can estimate the average delay incurred by the overall traffic within $X$~\cite{Kompella09}.

On the other hand, there isn't sufficient information for computing delay quantiles for domain $X$, i.e.,
we cannot make statements of the form ``$90$\% of the packets incurred delay below $10$msec within $X$.''
The only technique that we are aware of for computing delay quantiles for a domain requires knowing the delay incurred
by \emph{individual} packets within that domain~\cite{Sommers07}. Intuitively, this makes sense: An extreme example of a delay quantile
is the maximum delay incurred by a packet aggregate within $X$. Unlike average delay, maximum delay
cannot be computed without collecting per-packet information at the entrance and exit of $X$.

Now let's assume that there \emph{is} packet reordering within domain $X$.
In this case, the receipt collector cannot even compute the loss and average delay incurred within $X$,
because there is no guarantee that \HOPs 4 and 5 will break observed traffic into the same aggregates.

\subsection{Recap}

A simple protocol (like the strawman), where each domain produces receipts for each packet it receives and delivers, 
provides sufficient information for computing and verifying each domain's loss/delay performance;
however, the amount of resources required to store, process, and report per-packet state is (significantly) more
than a typical domain can afford today.
An aggregation-based protocol (like Difference Aggregator ++), where each domain produces per-aggregate receipts,
introduces tunable cost, but is susceptible to packet reordering and does not provide sufficient information for estimating
delay quantiles---only for computing loss and estimating average delay.
Finally, a sampling-based protocol (like Trajectory Sampling ++), where each domain produces receipts for sampled packets,
does provide sufficient information for estimating loss and delay quantiles \emph{and} introduces tunable cost,
yet is susceptible to sampling bias.

\subsection{\Sys Overview}

\Sys employs both sampling and aggregation---sampling to provide probabilistic delay-quantile measurements
and aggregation to provide precise loss measurements.

\Sys's sampling component shares elements with Trajectory Sampling ++ (\HOPs produce receipts
for a subset of observed packets and choose which packets to sample using hash functions), 
but prevents sampling bias in the following way.
The sampling function is keyed using \emph{future} traffic, making the samples unpredictable.
Specifically, a domain does not know whether it will have to report measurements on a particular packet until after 
it has forwarded that packet to its downstream neighbor. As a result, an unscrupulous
domain has no way to decide whether to ``sugarcoat'' its performance by
preferentially treating particular packets.

\Sys's aggregation component shares elements with Difference Aggregator ++ (\HOPs produce receipts
for packet aggregates and choose where to break each aggregate using hash functions), 
but provides accurate statistics in the face of packet reordering.
This is achieved by providing, on top of per-aggregate receipts, extra per-packet information
for a small window around the cutting points between packet aggregates.

One could ask, why use \emph{both} sampling and aggregation? After all, using sampling we can estimate both loss and
delay quantiles (provided we fix the sample bias issue), so why use aggregation at all?
One reason is that aggregation provides precise (as opposed to probabilistic) loss measurements and, as we will
see, once we have deployed the sampling component, the incremental cost of adding the aggregation component is trivial.
Another reason is to add extensibility to our mechanism. Even though we do not consider this scenario in this paper, 
``bad'' ISP behavior may consist not only of introducing loss and unpredictable delay, but also of modifying traffic;
the only way to detect such behavior is to use a content-processing technique like the one proposed 
in~\cite{Goldberg08}, which could be easily incorporated in our aggregation component, 
but not in a sampling-only mechanism.

\section{Voluntary Reporting}
\label{sec:reporting}

In this section, we describe what kind of information \Sys domains produce
and how that information is used to estimate and verify their performance.
We do not worry about \emph{how} this information is generated---we defer that to
the next two sections.

\paragraph*{Traffic Receipts}
Each \Sys \HOP generates receipts for the traffic it observes.
There are two kinds of receipts: 
\begin{enumerate}
\item A receipt for a set of sampled packets has form \\
$\receipt = \langle\pathID, \mathit{Samples}\rangle$.
\item A receipt for a packet aggregate has form \\
$\receipt = \langle\pathID, \aggID, \pktCount\rangle$.
\end{enumerate}

$\pathID$ specifies the \HOP path to which the corresponding sampled packets or packet aggregate belongs.
It has form $\langle \headerSpec, \previousHOPID, \nextHOPID, \checkValue\rangle$.
$\headerSpec$ specifies which part of a packet's headers is used to identify the packet's path;
it includes at least a source and destination origin-prefix pair.
$\previousHOPID$ and $\nextHOPID$ specify the previous and next \HOPs on this path.
$\checkValue$ is a value agreed upon between the reporting \HOP and
the \HOP that is at the other end of the same inter-domain link (e.g.,
\HOPs 3 and 4 in Figure~\ref{fig:sla}).  It is meant to lower-bound the
difference in timestamps one should expect between the two \HOPs.

$\mathit{Samples}$ is a sequence of $\langle \packetID,
\timestamp\rangle$ records, each corresponding to a single sampled measurement.
The packet identifier $\packetID$ is a digest of the packet's headers. 
$\timestamp$ specifies when the corresponding packet was observed at the \HOP.
The aggregate identifier $\aggID$ consists of the packet IDs 
of the first and last packets of the aggregate.  
$\pktCount$ is the number of packets observed by the \HOP within this aggregate. 

Upon receiving a packet, each \HOP classifies it into a \HOP path and an aggregate, counts
it against that aggregate's packet count, and decides whether to sample it.
Periodically, the \HOP generates traffic receipts for all the sampled packets \emph{and} 
aggregates it has observed since the last reporting time, which it disseminates
to all domains that observed the corresponding traffic.

\paragraph*{Receipt-based Statistics}
Consider \HOPs 4 and 5 in Figure~\ref{fig:sla} and suppose we collect all their receipts.
We now describe the types of statistics we can compute from these receipts.

Suppose \HOPs 4 and 5 use the same sampling algorithm, 
i.e., if one \HOP samples a packet $\packet$, the other \HOP also samples $\packet$
(provided $\packet$ is not lost before reaching the \HOP).
If the two \HOPs generate for $\packet$ receipts $\receipt_4^p$ and $\receipt_5^p$, respectively, 
then the packet's delay through $X$ was
$\receipt_5^p.\timestamp - \receipt_4^p.\timestamp$.
By computing the delay experienced by the sampled packets within $X$,
we can estimate upper and lower bounds for the delay experienced 
by all packets within $X$~\cite{Sommers07}. 

Now suppose \HOPs 4 and 5 use the same aggregation algorithm.
If the two \HOPs generate for the
same packet aggregate $\agg$ receipts $\receipt_4^\agg$ and
$\receipt_5^\agg$, respectively, then $X$ lost 
$\receipt_4^\agg.\pktCount - \receipt_5^\agg.\pktCount$ packets of the aggregate.

\paragraph*{Receipt Combination}
Receipts of either kind can be combined with others from the same \HOP to generate receipts
of a larger sample set or coarser aggregate.  For sampling receipts
combination is straightforward:
$$\uplus_i \receipt_i = \left\langle \pathID, \bigcup_i \mathit{Samples}_i \right\rangle$$
For aggregate receipts, consider $N$ consecutive aggregates, $\agg_i,
i=1..N$, from the same path, and the $N$ receipts, $\receipt^{\agg_i} =
\langle \pathID, \aggID_i, \pktCount_i\rangle$, produced for these
aggregates by a single \HOP.  We define the combination of these
receipts as
$$\uplus_i \receipt_i = \left\langle \pathID, \aggID, \sum_i \pktCount_i\right\rangle$$
where $\aggID$ is the identifier (first and last packet digest) 
of the union of all $N$ aggregates.

\paragraph*{Receipt Consistency}
Consider two receipts, $\receipt_5^p$ and $\receipt_6^p$, for the same sampled packet $\packet$, produced
by two \HOPs on opposite ends of the same inter-domain link (e.g., \HOPs 5 and 6, in Figure~\ref{fig:sla}).
The two receipts are considered \emph{consistent} with each other when all of the following hold:
\begin{eqnarray}
\receipt_5^p.\pathID.\checkValue & = & \receipt_6^p.\pathID.\checkValue\\
\receipt_6^p.\timestamp - \receipt_5^p.\timestamp & \leq & \receipt_5^p.\pathID.\checkValue
\end{eqnarray}
These rules express the fact that a correct inter-domain link does not introduce unpredictable delay:
the time at which a sampled packet is delivered by one \HOP and received by the other should differ
at most by a predictable $\checkValue$, set during configuration of that
link by the two involved domains.

Now consider two receipts, $\receipt_5^\agg$ and $\receipt_6^\agg$, for the same packet 
aggregate $\agg$, produced by two \HOPs on opposite ends of the same inter-domain link.
The two receipts are considered \emph{consistent} with each other when:
$$\receipt_5.\pktCount = \receipt_6.\pktCount$$
This rule represents the fact that a correct inter-domain link does not introduce packet
loss---hence, the number of packets delivered by one \HOP and received by the other should be the same.

If a receipt collector gets inconsistent receipts from two neighbors, it discards both receipts and notifies both neighbors of the
inconsistency, such that the liar is exposed to the neighbor it implicated, as in the strawman (\S\ref{sec:new:strawman}).

\paragraph*{(No) Clock Synchronization}
\Sys does not require that \HOPs have synchronized clocks.
However, it is to a participating domain's best interest to keep its \HOPs
reasonably synchronized (e.g., at the granularity of a millisecond, achievable with 
NTP~\cite{Mills06}), since its delay performance will be estimated based on the 
timestamps reported by different \HOPs.
Moreover, it is to two neighboring domains' best interest to keep adjacent \HOPs
(like 3 and 4 in Figure~\ref{fig:sla}) reasonably synchronized, otherwise their
timestamp difference will exceed the reported $\checkValue$ and the two neighbors
will generate inconsistent receipts (hence appear to have a problematic inter-domain
link or be involved in a lie).

We should note that domains are free to report arbitrarily large $\checkValue$ values:
nothing prevents \HOPs 3 and 4 from keeping de-synchronized clocks and reporting a 
$\checkValue$ of several seconds between them.
That, however, does make it look like they are connected through an awfully slow inter-domain 
link---not a good feature to advertise to their customers and peers.

\section{Bias-resistant, Tunable Sampling}
\label{sec:sampling}

We now describe how each \HOP chooses which packets to sample.
Our sampling algorithm prevents domains from exaggerating their performance by biasing their samples
(\S\ref{sec:sampling:bias}), while it maximizes the number of packets that are commonly sampled by 
all \HOPs that observe them, while allowing each \HOP to choose its own sampling rate (\S\ref{sec:sampling:tunability}),
even in the face of loss and packet reordering (\S\ref{sec:sampling:loss}).

\subsection{Bias Resistance}
\label{sec:sampling:bias}

Instead of sampling packets in real time, each \HOP maintains state on
\emph{all} observed packets, but only for a fixed, \emph{short} period of
time (ten milliseconds or so).  After that period of time has elapsed, the \HOP is told which
of the stored per-packet state to keep and which to discard.  Since an
ISP learns whether a packet's fate will affect estimates of its
performance only \emph{after} it has forwarded that packet, it cannot treat sampled packets preferentially.

A dishonest \HOP could, in theory, store every single packet, wait to learn whether the packet has to be sampled,
\emph{then} decide how to treat the packet.
However, that means delaying all traffic at the \HOP by ten milliseconds or so (an order of magnitude above
the delay introduced by a correctly functional router)---not to mention that it requires buffering ten milliseconds'
worth of traffic, which, for a $10$Gbps interface would require $25$MB (i.e., several chips) of expensive SRAM storage.

A key question is \emph{who} tells each \HOP which packets to delay-sample.
A na\"ive approach would be to use explicit signaling; for example,
in Figure~\ref{fig:sla}, domain $S$ could explicitly tell all \HOPs in path $\dlp$
which packets to sample from each aggregate sent from $S$ to $D$ along $\dlp$.
That, however, would essentially require every source domain to set up virtual circuits
along all Internet paths that observe its traffic.
Instead, each \HOP decides whether to delay-sample a packet
based on the \emph{contents of another packet} sent \emph{later} on the same path.
In this sense, domain $S$ implicitly dictates which of its
packets should be sampled, through the traffic it subsequently routes
via $\dlp$
anyway.

\begin{algorithm}[t]
\caption{\label{alg:dsamp} $\dsamp(\packet, \marker, \samplingthresh)$}
\begin{tabular}{l c l}
Input & $\packet$ & // new packet \\
Input & $\marker$ & // marker threshold \\
Input & $\samplingthresh$ & // sampling threshold \\
Initially & $\mathit{TempBuffer} \leftarrow \emptyset$ & // packet buffer \\
Initially & $\receipt \leftarrow \emptyset$ & // current receipt \\
\end{tabular}
\begin{algorithmic}[1]
\IF {$\descriptor(\packet) > \marker$}
\FORALL {packets $q$ in $\mathit{TempBuffer}$}
\IF {$\samplingFunction(\descriptor(q), \descriptor(\packet)) > \samplingthresh$}
\STATE {Add $\langle\descriptor(q), \timestamp(q)\rangle$ to $\receipt.\samples$}
\ENDIF
\ENDFOR
\STATE {Empty $\mathit{TempBuffer}$}
\STATE {Add $\langle\descriptor(p), \timestamp(p)\rangle$ to $\receipt.\samples$}
\ELSE
\STATE {Add $\descriptor(\packet)$ to $\mathit{TempBuffer}$}
\ENDIF
\end{algorithmic}
\end{algorithm}

Algorithm~\ref{alg:dsamp} shows what happens when a \HOP observes a new
packet $\packet$ from path $\dlp$; the algorithm assumes that the \HOP
maintains a temporary buffer with per-packet state for all the packets
observed from $\dlp$.  If the packet satisfies a certain condition, it
is chosen as a ``marker'' packet (line $1$).  In that case, its contents
determine which of the already observed packets to sample (lines
$2$--$4$) discarding the rest (line $5$).  The marker packet itself is also
sampled (line $6$).
Observe that \HOPs maintain state for all packets only during the short period 
of time until the next marker packet is observed.

The \emph{marker value} $\marker$, which determines which packets are
``markers,'' is a system-wide constant specified by \Sys at design
time; when there is no loss, all \HOPs in $\dlp$ select the same packets as markers. 
In contrast, the \emph{sampling threshold} $\samplingthresh$, which determines which 
packets are sampled, is a local parameter, chosen independently at each \HOP.  
If all \HOPs in $\dlp$ choose the same $\samplingthresh$, they all sample 
the same packets (modulo the packets that are lost). We turn next to what 
happens when different \HOPs select different sampling thresholds.

\subsection{Tunability}
\label{sec:sampling:tunability}

Each \HOP chooses its own sampling rate.
At the same time, given $N$ \HOPs observing the same packet sequence 
and their sampling rates, we maximize the number of packets that are commonly sampled by all \HOPs.

The key element that enables this property is the inequality in line 3 of Algorithm~\ref{alg:dsamp}:
Consider \HOPs $1$ and $2$, with sampling thresholds
$\samplingthresh_1$ and $\samplingthresh_2 < \samplingthresh_1$.
Suppose that $p$ is a packet sampled by \HOP $1$ and $q$ is the first marker packet
observed after $p$ by \HOP $1$.
Since \HOP $1$ samples $p$, this necessarily means that 
$\samplingFunction(\descriptor(q), \descriptor(p)) > \samplingthresh_1 > \samplingthresh_2$,
which means that \HOP $2$ also samples $p$; hence,
\HOP $2$ samples \emph{at least} all packets sampled by \HOP $1$.
So, even though each \HOP chooses its sampling rate independently,
if there is no packet loss or reordering, different \HOPs never sample partially overlapping packet sets.
%For example, if \HOPs $1$, $2$, and $3$ from Figure~\ref{fig:sla} have sampling thresholds
%$\samplingthresh_1 > \samplingthresh_2 > \samplingthresh_3$, they
%may respectively sample sets $\{p_1\}$, $\{p_1, p_5\}$, and $\{p_1, p_3, p_5, p_7\}$,
%or $\{p_2\}$, $\{p_2, p_6\}$, and $\{p_2, p_4, p_6, p_8\}$,
%but not $\{p_1\}$, $\{p_2, p_5\}$, and $\{p_2, p_4, p_6, p_8\}$.

\subsection{Sampling Under Loss and Reordering}
\label{sec:sampling:loss}

Loss and reordering decrease the number of commonly sampled packets.
E.g., if a marker packet get lost between two \HOPs, it causes them to sample
arbitrarily different packet sets for several milliseconds---until the next marker arrives.
The good news is that it takes unlikely amounts of (non-purposeful) loss/reordering to 
significantly impact the estimation accuracy of the mechanism. For instance, in Section~\ref{sec:eval},
we show that, if \HOPs 4 and 5 sample $1$\% of the observed traffic, and the link between them experiences
$25$\% packet loss, the delay between the two \HOPs can still be estimated with an accuracy of $2$msec. 
This accuracy is sufficient for verifying today's SLAs, which typically promise 
intra-domain delays on the order of multiple tens of milliseconds~\cite{SprintSla}.

\iffalse
We illustrate the second case with an example:
Suppose \HOPs $4$ and $5$ from Figure~\ref{fig:sla} have the same sampling threshold $\samplingthresh$.
\HOP $4$ observes packet sequence $\langle p_1, p_2, q_1, p_3, p_4, q_2 \rangle$,
where $q_1$ and $q_2$ are marker packets; it samples $p_1$ 
(because $\samplingFunction(\descriptor(q_1), \descriptor(p_1)) > \samplingthresh$)
and $p_3$ (because $\samplingFunction(\descriptor(q_2), \descriptor(p_3)) > \samplingthresh$).
Suppose $q_2$ is dropped, such that \HOP $5$ observes packet sequence 
$\langle p_1, p_2, p_3, p_4, q_2 \rangle$; it samples $p_2$
(because $\samplingFunction(\descriptor(q_2), \descriptor(p_2)) > \samplingthresh$) and $p_3$.
So, even though the two \HOPs have the same sampling threshold and should have sampled
the same packets, they sampled different packets because marker $q_1$ was lost.

Reordering can have a similar effect: if \HOP $4$ observes packet sequence
$\langle p_1, p_2, q_1, p_3, p_4, q_2 \rangle$, but \HOP $5$ observes
$\langle q_1, p_1, p_2, p_3, p_4, q_2 \rangle$, the outcome will be the same
as if marker $q_1$ was lost. However, we should note that, in practice,
only packets that are close to one another can be reordered, while
marker packets are separated by thousands of non-marker packets; hence,
few sampled packets can be actually affected by marker reordering (more on this
in Section~\ref{sec:eval}).
\fi

An under-performing domain (say $X$ in Figure~\ref{fig:sla}) could drop all marker packets, 
causing the next domain ($N$ in our example) to sample all the wrong packets; this would 
ensure that $X$'s performance is never verified according to $N$'s receipts.
First, note that such behavior from $X$ is detrimental to $N$ (because it prevents it 
from producing correct receipts), hence $N$ has a clear incentive to expose and stop it.
Second, such behavior is bound to be exposed,
because marker packets are expected to be always sampled and reported on:
if $X$ drops a marker $q$, it either has to admit dropping it or lie and be inconsistent with $N$'s
claim that it never received $q$; either way, if $X$ consistently drops markers, it is either
globally exposed as misbehaving or locally exposed as such to $N$.

\section{Tunable Aggregation}
\label{sec:aggregation}

We now describe how each \HOP chooses which packets to assign to the same aggregate.
Like our sampling, our aggregation is ``tunable,'' i.e.,
we allow each \HOP to choose its own degree of aggregation, according to the locally available resources.
This raises the following challenge: when \HOPs aggregate differently, they
produce receipts on different aggregates; how can one combine such receipts 
to estimate domain performance and perform consistency checking?
We first describe this challenge in more detail (\S\ref{sec:tunable:problem}),
then present our solution in two parts---first assuming no loss or reordering 
(\S\ref{sec:tunable:basic}), then removing this assumption
(\S\ref{sec:tunable:reordering}).

\smallskip
\noindent
{\bf Terminology and Notation}:
We borrow the following terminology and notation from set theory
(illustrated through the examples of Table~\ref{tab:aggExample}):
\begin{CompactEnumerate}
\item 
A \emph{partition} of a packet set $\stream$ is a set of non-overlapping aggregates
whose union is equal to $\stream$.
Given a partition $\aggset$ of some packet set,
each packet that is the first packet of an aggregate in $\aggset$
is called a \emph{cutting point}.
For example, $\packet_1$ and $\packet_3$ are cutting points
in $\aggset = \{\{\packet_1, \packet_2\}, \{\packet_3, \packet_4\}\}$.

\item
Suppose $\aggset_1$ and $\aggset_2$ are partitions of the same packet set.
We say that $\aggset_1$ is \emph{coarser} than $\aggset_2$
(or $\aggset_2$ is \emph{finer} than $\aggset_1$), denoted by $\aggset_1 \ge \aggset_2$,
when each aggregate in $\aggset_1$ is a union of aggregates in $\aggset_2$.

More formally, we say that $\aggset_1 \ge \aggset_2$, when: \\
$\exists \{\aggb_i|\aggb_i \in \aggset_2\}: \bigcup_i\aggb_i = \agg , \forall \agg \in \aggset_1$.

\item
Suppose $\aggset_i, i = 1..N$, is a partition of packet set $\stream$.
We say that $\aggsetj$ is the \emph{join} of $\aggset_1, \aggset_2, ...\aggset_N$,
denoted by $\aggsetj = \join(\aggset_1, \aggset_2, ...\aggset_N)$,
when $\aggsetj$ is the finest partition of $\stream$ that is coarser than all $\aggset_i$.

More formally, we say that \\
$\aggsetj = \join(\aggset_1, \aggset_2, ...\aggset_N)$,
when: \\
$\aggsetj \le \aggsetj' \; \forall \aggsetj':  \; \aggsetj' \ge \aggset_i \; \forall \; i$, \\
where $\aggsetj'$ is also a partition of $\stream$.
\end{CompactEnumerate}

\begin{table}[t] 
\centering
\small
\begin{tabular}{l|l} 
$\aggset_1 = 
\{\{\packet_1\}, \{\packet_2\}, \{\packet_3\}, \{\packet_4\} \}$ 
& \\
$\aggset_2 = 
\{ \{\packet_1, \packet_2\}, \{\packet_3, \packet_4\} \} \ge \aggset_1$
& $\join(\aggset_1, \aggset_2) = \aggset_2$ \\
$\aggset_3 = 
\{\{\packet_1\}, \{\packet_2, \packet_3\}, \{\packet_4\}\} \ge \aggset_1$
& $\join(\aggset_2, \aggset_3) = \aggset_4$ \\
$\aggset_3' = 
\{\{\packet_1\}, \{\packet_2\}, \{\packet_3, \packet_4\}\} \ge \aggset_2$
& $\join(\aggset_2, \aggset_3') = \aggset_2$\\
$\aggset_4 = 
\{ \{\packet_1, \packet_2, \packet_3, \packet_4\} \} \ge \aggset_2, \aggset_3$  
& 
\end{tabular}
\caption{\label{tab:aggExample}
\small
Different partitions of packet set $\stream = \{ \packet_1, \packet_2, \packet_3, \packet_4 \}$
and some join examples. Note that not all partitions of $\stream$ have
a ``$\ge$'' relationship, e.g., we cannot say that $\aggset_2 \ge \aggset_3$
nor that $\aggset_3 \ge \aggset_2$.}
\end{table}

\subsection{The Partitioning Problem}
\label{sec:tunable:problem}

If we view all traffic sent on path $\dlp$ as a packet set $\stream$,
then we can say that each \HOP in $\dlp$ that performs packet aggregation \emph{computes a partition} of $\stream$.

When two \HOPs produce different aggregate sets from the same packet set,
a domain that collects their receipts cannot directly perform consistency checking
as described in Section~\ref{sec:reporting}.
However, it can try to find traffic receipts from one \HOP that, when combined, 
exactly correspond to traffic receipts (and aggregates) from the other \HOP, and 
then proceed with the calculations and verification from Section~\ref{sec:reporting}.
This corresponds to computing the join of the two aggregate sets as defined above 
to find the finest aggregate set over which statistics can be computed across the
receipts from the two \HOPs.

For instance, suppose two \HOPs observe packet set $\stream$ from Table~\ref{tab:aggExample} 
and, respectively, produce aggregate sets $\aggset_2$ and $\aggset_3$ (from the same table).
A domain that collects their receipts can combine each \HOP's receipts and
produce the receipt that the \HOP would have produced for the (single) aggregate in aggregate set $\aggset_4$.
So, the two \HOPs' claims can be checked for consistency only with respect to the aggregates in
the coarser aggregate set $\join(\aggset_2, \aggset_3) = \aggset_4$.

Although this approach is general---there is always a join of two
aggregate sets over which a verifier can compute \emph{some} combined
receipts and, therefore, \emph{some} performance statistics---the
quality of the results varies.  Intuitively, we would want the join of
fine-grained aggregate sets to be just as fine-grained; otherwise
information obtained and forwarded at high resource cost would end up
lost in translation.  In the example above, the join of $\aggset_2$ and
$\aggset_3$ is $\aggset_4$, a single-aggregate aggregate set, even
though the input aggregate sets and traffic receipts afforded multiple
data points each from either \HOP.  In contrast, an equally ``expensive''
aggregate set $\aggset_3'$ from the second \HOP, would have allowed the verifier to
compare receipts on $\join(\aggset_2, \aggset_3') = \aggset_2$, which conserves all
information from the first \HOP and only combines two of the three receipts from
the second one.

Our goal then is:
\emph{to design a partitioning algorithm that results in the
finest possible join given the rate at which each \HOP can produce new aggregates}.

\subsection{Basic Solution}
\label{sec:tunable:basic}

At a high level, \Sys limits domains' choice of packet aggregation so
as to produce ``good'' aggregates with respect to join and combination,
while allowing them to tune how fine their choice is.  

Algorithm~\ref{alg:dpart} shows what happens when a \HOP observes a new packet $p$ from path $\dlp$;
the algorithm assumes that the \HOP maintains one ``open'' receipt per path.
If the packet's contents satisfy a certain condition (line $1$), then the current aggregate 
for path $\dlp$ is closed (line $2$) and the packet is classified in a new aggregate (line $4$); 
otherwise, the packet is classified in the current aggregate (line $5$).
Observe that this algorithm requires constant state per aggregate and constant computation per 
packet (i.e., its state size and per-packet computation are not proportional to aggregate size).

Algorithm~\ref{alg:dpart} ensures that \HOP $2$ with partition
threshold $\descthresh_2$ will partition a stream at least at the same points as
\HOP $1$ with partition threshold $\descthresh_1 > \descthresh_2$.
So, even though each \HOP chooses its partitioning rate independently,
if there is no loss or reordering, different \HOPs never produce partially
overlapping aggregate sets.
For instance, if \HOPs $1$ and $2$ from Figure~\ref{fig:sla} observe packet sequence
$\langle p_1, p_2, ... p_8 \rangle$ and have partition thresholds $\descthresh_1 > \descthresh_2$,
they may respectively produce aggregate sets 
$\{\{p_1, p_2, p_3, p_4\}, \{p_5, p_6, p_7, p_8\}\}$ and
$\{\{p_1, p_2\}, \{p_3, p_4\}, \{p_5, p_6\}, \{p_7, p_8\}\}$, 
but not
$\{\{p_1, p_2, p_3, p_4\}, \{p_5, p_6, p_7, p_8\}\}$ and
$\{\{p_1\}, \{p_2, p_3\}, \{p_4, p_5\}, \{p_6, p_7\}, \{p_8\}\}$.

\begin{algorithm}[t]
\caption{\label{alg:dpart}$\mathit{Partition}(\packet, \descthresh)$} 
\begin{tabular}{l c l}
Input& $\packet$ & // new packet \\
Input& $\descthresh$ & // partition threshold \\
Initially& $\receipt = \emptyset$ & // current receipt\\
%Output& $\langle \dlp, \receipt \rangle$ & // stream of receipts for path $\dlp$
\end{tabular}
\begin{algorithmic}[1]
\IF {$\descriptor(\packet) > \descthresh$}
%\COMMENT{Last packet ID makes part of the current aggregate ID.}
%\STATE $\receipt.\checkValue \leftarrow \ldots$ \COMMENT{Constant from
%  configuration file}
\STATE Close receipt $\receipt$ for aggregate $\receipt.\aggID$
\STATE Open new receipt $\receipt \leftarrow \emptyset$
\STATE $\receipt.\aggID.\firstPacket \leftarrow \packet$ 
%\COMMENT{First packet ID makes part of the next aggregate ID.}
\ENDIF
\STATE $\receipt.\aggID.\lastPacket \leftarrow \packet$ 
\STATE $\receipt.\pktCount \leftarrow \receipt.\pktCount + 1$
\end{algorithmic}
\end{algorithm}

\subsection{\hspace{-0.2cm} Partitioning Under Loss and Reordering}
\label{sec:tunable:reordering}

Loss can decrease the fine-ness of the join of the produced aggregate sets:
Suppose \HOPs $1$ and $2$ produce aggregate sets 
$\{\{p_1, p_2, p_3, p_4\}, \{p_5, p_6, p_7, p_8\}\}$ and
$\{\{p_1, p_2\}, \{p_3, p_4\}, \{p_5, p_6\}, \{p_7, p_8\}\}$;
the join of the two sets is $\{\{p_1, p_2, p_3, p_4\}, \{p_5, p_6, p_7, p_8\}\}$ 
(the coarsest of the two aggregate sets).
However, if $p_5$ is lost before \HOP $2$, then the latter produces aggregate set 
$\{\{p_1, p_2\}, \{p_3, p_4, p_5, p_6\}, \{p_7, p_8\}\}$; now, the join of the 
two sets is $\{\{p_1, p_2, p_3, p_4, p_5, p_6, p_7, p_8\}\}$ (the worst
possible in this example).
So, loss can cause a combination of aggregates that would otherwise have been split 
using the lost packet as a cutting point, which, in turn, reduces the fine-ness of the join. 

The good news is that, although loss does decrease the fine-ness of the resulting join,
the degradation is smooth, because the probability of coarsening the granularity of a 
measurement is conditioned on a cutting point being lost, not on arbitrary packet loss and, 
even then, not all cutting points can cause a violation of the total order when lost. 
For instance, in Section~\ref{sec:eval}, we show that, if \HOPs 4 and 5 generate an
aggregate receipt for every $100,000$ packets, and the link between them experiences
$25$\% loss, the loss between the two \HOPs can still be computed for every
$150,000$ packets, on average. 
Note that being able to compute domain loss at such granularity is more than sufficient
for verifying today's SLAs, which typically promise a certain level of packet loss
\emph{per month} (a duration that corresponds to billions of packets,
assuming a traffic rate of a few tens of Mbps along each path)~\cite{SprintSla}.

Reordering can also decrease the fine-ness of the join of the produced aggregate sets:
Consider path $\dlp$ from Figure~\ref{fig:sla} and original packet
sequence $\seq = \langle \packet_1, \packet_2, ... \packet_8 \rangle$
sent along $\dlp$.  Suppose \HOP $1$ observes this sequence and
partitions it into aggregate set $\aggset = \{\{\packet_1, \packet_2,
\packet_3, \packet_4\}, \{\packet_5, \packet_6, \packet_7,
\packet_8\}\}$.  \HOP $4$ observes sequence $\langle \packet_1,
\packet_2, \packet_3, \packet_5, \packet_4, \packet_6, \packet_7,
\packet_8 \rangle$ due to reordering somewhere between the two \HOPs. Even though it uses the same
algorithm, it partitions the sequence into $\aggset' = \{ \{\packet_1,
\packet_2, \packet_3\}, \{\packet_5, \packet_4, \packet_6, \packet_7,
\packet_8\} \}$.  The two aggregate sets are not ordered according to
the ``finer than'' relation, so their join is the entire sequence, an
undesirable effect of reordering.

In practice, packets are reordered only when they are transmitted close
to one another (according to the most recent Internet-wide experiment we are aware of,
packets transmitted more than half a millisecond apart were not reordered~\cite{Gharai04}).  
Hence, we define, for each path $\dlp$, a \emph{safety inter-arrival threshold} $\maxJitter$ 
and assume that two packets that follow $\dlp$ can be reordered only if they are observed 
(at any \HOP) less than $\maxJitter$ time units away from one another.
This assumption allows us to bound the coarseness of the join at the
cost of keeping extra per-aggregate state.

At a high level, we alter the mechanism of Algorithm~\ref{alg:dpart}
to add \emph{patch up} information in every receipt. A verifier can use
this patch up information to make ``misaligned'' receipts from different
\HOPs align better, thereby enabling a better join of the corresponding
aggregate sets and consequently better-quality traffic statistics. 

More specifically, a traffic receipt for a packet aggregate also specifies
the sequence of packets observed $J$ time units around the cutting point. 
In the above example, \HOP 1 reports sequence $\langle \packet_3, \packet_4,
\packet_5, \packet_6 \rangle$ in its receipt for the
first aggregate, and \HOP 4 reports sequence $\langle \packet_2, \packet_3,
\packet_5, \packet_4 \rangle$ in its receipt for the first aggregate.
In general, a receipt is extended from the earlier definition to be
$\langle\pathID, \aggID, \pktCount, \aggTrans \rangle$,
where $\aggTrans$ is the sequence of packet identifiers that correspond to
the packets observed within a window of $2\maxJitter$ from the aggregate's last
packet.

Using this information, the verifier can transform one \HOP's receipts
to match what the \HOP would have generated, had it observed the same
packet sequence with another \HOP.  In our particular example, \HOP 1
reports observing packet $\packet_4$ before cutting point $p_5$, while
\HOP 4 reports observing it after the cutting point.  Consequently, the
verifier would transform \HOP 4's receipts by ``migrating'' $\packet_4$
from the later to the earlier aggregate (i.e., decrementing the packet
count of the former and incrementing the packet count of the
latter). With this transformation, \HOP 4's receipts correspond to the
same aggregates with \HOP 1's receipts, hence the verifier can proceed
with the performance computation and verification of
Section~\ref{sec:reporting}.  
%This transformation process 
%works not only for packet counts but
%also for any other commutative and associative aggregate measure.

If adding per-packet state to aggregate receipts sounds like too much overhead, 
take into account that a \HOP is supposed
to \emph{choose} how many packets to assign to each aggregate according to its resources.
E.g., a \HOP may choose to cover minutes' worth of traffic with each aggregate; 
in this case, including in each per-aggregate receipt per-packet state for the few 
packets observed around the end of the aggregate is significantly less
expensive than maintaining per-packet state. We quantify this per-packet overhead
in Section~\ref{sec:eval}.

\section{Evaluation}
\label{sec:eval}

We now compute the resource overhead incurred by \Sys domains,
and quantify the quality with which each domain's performance is estimated.

We consider the case where \HOP functionality is implemented in border routers, 
as part of a NetFlow-like monitoring platform that operates partly in the router's 
data-plane and partly in its control plane.
The data-plane part handles per-packet operations and collects per-aggregate state in a monitoring cache;
we refer to it as the \emph{collector module}.
The control-plane part periodically reads the state from the data-plane and performs further processing;
we refer to it as the \emph{processor module}.

As a proof of concept, we implemented the collector and processor modules in Click
(although, in a real router, the former would be implemented in hardware,
close to the router's forwarding plane, e.g., as part of a NetFlow engine).
Our implementation uses the ``Bob'' hash function (because it has been shown to work
well with Internet traffic~\cite{Molina05}) to compute packet digests
and applies it to each packet's IP and transport headers.
The collector's monitoring cache is updated from traffic traces (as opposed to actual
network traffic).  We used traces from a Tier-1 ISP, provided by CAIDA.

\subsection{Overhead}

\paragraph*{Memory and Processing}
The amount of memory and processing resources needed for the processor module is tunable.
The processing module reads receipts from the monitoring cache and prepares them for storage or dissemination.
The rate at which new receipts appear in the monitoring cache (hence need to be read and processed)
depends directly on the locally chosen sampling and partition thresholds. Hence, a domain can directly control
the amount of memory and processing cycles spent by the processing module by varying these two thresholds 
(a demonstration of the resulting trade-off follows).

The collector module maintains state for each ``active path,'' i.e., each source-destination
origin-prefix pair that is currently sending traffic through the specific \HOP; this per-path
state consists at least of one ``open'' aggregate receipt 
(a $\pathID$, $\aggID$, and $\pktCount$---roughly $20$ bytes).
E.g., if a \HOP observes traffic from $100,000$ paths at the same time,
it needs a $2$MB monitoring cache.

Moreover, the collector module maintains a temporary packet buffer,
where it stores $\langle \packetID, \timestamp \rangle$ pairs ($4$ and $3$ bytes, respectively)
for all packets observed within $\maxJitter$ time units.
At first, this seems to be cause for concern---what happens with high-rate paths
that observe millions of packets per second?
In reality, however, the per-packet state that needs to be kept is modest:
Recall that $\maxJitter$ is our ``safety threshold''---when two packets are observed
more than $\maxJitter$ time units apart, we assume that they cannot be reordered.
A conservative choice is to set $\maxJitter$ to $10$msec---an order of magnitude above the millisecond
threshold that we need according to the latest Internet reordering measurements we are aware of~\cite{Gharai04}.
An OC-192 interface observes at most $10$Gbps. If we assume an average packet size of $400$B,
$10$Gbps corresponds to $3.125$Mpps per direction, which means that a \HOP would need a $436$KB temporary buffer for each
$10$Gbps interface.
Assuming an (implausible) worst-case traffic of all minimum-size packets, $10$Gbps correspond to $20$Mpps per direction,
which means that a \HOP would need a $2.8$MB temporary buffer for each $10$Gbps interface.
So, even assuming worst-case traffic, the amount of buffering we need fits into a single SRAM chip.

Finally, for each packet $\packet$, the collector looks up the packet's $\pathID$; computes 
$\descriptor(\packet)$ and a timestamp;
updates the corresponding $\pktCount$; and stores the digest and timestamp to the temporary packet buffer.
This amounts to three memory accesses, one hash function, and one timestamp computation per packet.
Moreover, whenever a marker packet is observed, the \HOP goes through the temporary packet buffer
and discards state for the packets that are not delay-sampled, which adds one more memory access per packet.
Such processing, though not currently supported by routers, is within the capabilities of modern
hardware and in line with the guidelines set by the IETF Packet Sampling group~\cite{psamp}.
 
\paragraph*{Bandwidth}
We have said that each domain makes each receipt available to every other domain that
observed the corresponding traffic.
Whether this happens pro-actively (through a constant receipt stream) or on-demand
(e.g., through a secure web interface), receipt dissemination introduces, in each path,
bandwidth overhead that depends on (1) the number of \HOPs on that path and (2) the rate
at which each of these \HOPs produces receipts. 

Again, this seems, at first, to be cause for concern---one could argue that introducing bandwidth 
overhead that grows with the total number of \HOPs per path is not a ``scalable'' approach.
In practice, this dependence on the number of \HOPs is not a problem:
Paths consist on average of $3$--$4$ domains, hence $4$--$6$ \HOPs (check the ``Average AS path length'' and
``Average address weighted AS path length'' entries in~\cite{potaroo}).
To be conservative, we consider a $10$-domain path, 
where each \HOP puts on average an ambitious $1000$ packets per aggregate and samples $1$\% of the path's packets.
Given receipt size ($22$ bytes),
this path will incur an overhead of $0.2$ bytes per packet; assuming $400$ bytes per packet,
this leads to a $0.046$\% bandwidth overhead for the path.

\paragraph*{Click Implementation}
As a proof of concept, we configured an eight-core Intel Nehalem server as a standard IPv4 router
and fed to it a real trace. Then we measured the router's performance with and without our 
\Sys modules loaded and saw no difference (in both cases, the server routed $25$Gbps).
This is not surprising, given that, when fed realistic traffic, a Nehalem server is bottlenecked at the I/O,
whereas our \Sys modules burden the CPU.

\subsection{Quality}

\paragraph*{Methodology}
We consider the case where domain $X$ from Figure~\ref{fig:sla} is congested, and $X$'s
delay performance is estimated from its receipts.
Each experiment consists of:
(1) extracting a packet sequence $\seq$ from one of our traces and consider the case where 
$\seq$ is sent through domain $X$;
(2) simulating a scenario where the intra-domain path between \HOPs 4 and 5 is congested;
(3) generating the receipts that $X$ would generate for packet sequence $\seq$;
(4) estimating $X$'s performance as a verifier would estimate it based on $X$'s receipts, i.e.,
using the technique from~\cite{Sommers07}.
(5) comparing that to $X$'s actual performance.

For step 1, we use traces provided by CAIDA, collected in 2008 from a Tier-1 ISP.
When we say that we ``extract a packet sequence'' from a trace, we mean that we 
extract all packets that carry a given source and destination origin-prefix pair.
The point of using real traces is to verify that our sampling and aggregation
algorithms work well given an actual packet stream---e.g., when a domain chooses its
sampling threshold so as to sample $1$\% of the observed traffic, it indeed
samples $1$\%. The results we show correspond to a particular packet sequence (of $100,000$ packets 
per second), but all traces and packet sequences we tried gave us consistent results.

For step 2, we ``introduce'' loss and delay in the chosen packet sequence.
To introduce loss, we discard a subset of the packets, chosen using the
Gilbert-Elliot loss model~\cite{Gilbert}.
Introducing delay is more complicated, as we are not aware of any commonly acceptable
delay model for Internet traffic.
Instead, we use the NS simulator to create realistic congestion scenarios, and generate the sequence
of delay  values that our packet sequence would encounter in each case.
We consider different congestion scenarios, where long-lived TCP or UDP flows compete for/saturate
the bandwidth of a bottleneck link, but show results only for the scenario that introduced the 
highest delay variance in the shortest time scale.

\iffalse
Regarding step 4: To estimate $X$'s delay performance, we use the algorithm proposed in~\cite{Sommers07}.
This algorithm takes as input the delay values experienced by the sampled packets within $X$ during a given
time period, and outputs lower and upper bounds for $X$'s delay cumulative distribution function (CDF) for that same period.
When the actual CDF values fall within the bounds, the estimation is correct.
The closer the lower and upper bounds are to one another, the more accurate the estimation.
Hence, we talk about the ``accuracy'' with which $X$'s delay performance is estimated,
we refer to the difference, in milliseconds, between these lower and upper bounds.
\fi

\begin{figure}[t]
\includegraphics[width=3.2in]{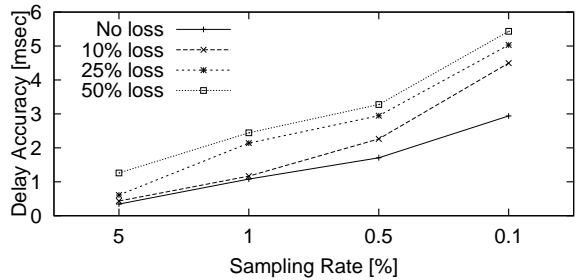}
\caption{\small 
The accuracy with which domain $X$'s delay performance is estimated
as a function of $X$'s sampling rate, for different levels of loss,
when $X$ uses our sampling algorithm.
%On the left: congestion is caused by $100$ long-lived TCP connections.
Congestion is caused by a bursty, high-rate UDP flow.
}
\label{fig:sampling}
\vspace{-0.3cm}
\end{figure}

\paragraph*{Accuracy of Estimated Delay}
By reducing its sampling rate, a \Sys domain can reduce the amount of resources it spends sampling,
at the cost of its delay performance being estimated with lower accuracy.
We now examine this trade-off.

We run a set of experiments where we vary domain $X$'s sampling rate.
Figure~\ref{fig:sampling} (consider the ``No loss'' curve) shows the accuracy with which 
$X$'s delay performance is estimated, as a function of the sampling rate.
We see that, reducing the sampling rate results in \emph{smooth} accuracy degradation.
Even if $X$ samples only $0.1$\% of the observed traffic, 
its delay performance is estimated with sub-millisecond accuracy.

Next, we examine how packet loss affects our sampling algorithm, hence the accuracy with
which a \Sys domain's delay performance is estimated.
We run a set of experiments where we vary both $X$'s sampling rate
and the amount of packet loss introduced by $X$.
Figure~\ref{fig:sampling} shows how accuracy degrades with lower sampling rate,
for different loss values.
We see that, when $X$ samples $1$\% of the observed traffic
and $25$\% of this traffic is lost within $X$, $X$'s delay performance is still estimated with
an accuracy of $2$ msec.
This robustness in the face of loss is partly due to our sampling algorithm 
and partly owed to the estimation algorithm from~\cite{Sommers07} (which works
well even with few samples).

\paragraph*{Granularity of Computed Loss}
We now examine how packet loss affects the granularity at which
a \Sys domain's loss performance is computed.

We run a set of experiments where we fix $X$'s aggregation rate (such that it produces one aggregate
every $100,000$ packets) and vary the amount of packet loss introduced by $X$. 
Figure~\ref{fig:sampling} shows the granularity at which $X$'s loss performance can be computed,
as a function of the loss rate.
We see that, when there is no loss, $X$'s loss performance can be computed over $1$sec periods
(because $X$ produces a new aggregate every $100,000$ packets, which, for the particular packet
sequence we are considering, corresponds to $1$ sec).
As the level of loss increases, granularity worsens---i.e., a verifier that collects $X$'s receipts
cannot always compute $X$'s loss performance over $1$sec periods.
However, the degradation is, again, smooth:
even if $X$ loses $25$\% of the observed traffic,
its loss performance is computable over periods of $1.5$sec.
This robustness in the face of loss is due to our aggregation algorithm,
which maximizes the number of common aggregates across \HOPs---essentially 
enables \HOPs not to fall ``out of sync'' when packets get lost.

\begin{figure}[t]
\includegraphics[width=3.2in]{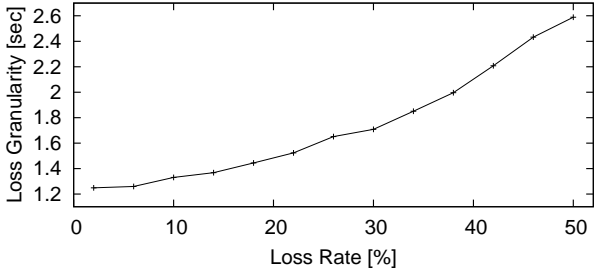}
\caption{\small
The granularity at which domain $X$'s loss performance is computed
as a function of the loss rate introduced by $X$,
when $X$ uses our aggregation algorithm.
}
\label{fig:loss}
\vspace{-0.3cm}
\end{figure}

\paragraph*{Verifiability}
We have demonstrated that a \Sys domain's loss and delay performance can be accurately estimated
from its receipts, even when the domain samples $1$\% of the observed traffic, puts hundreds of 
thousands of packets into a single aggregate, and is severely congested (to the point of losing more 
than $25$\% of the observed traffic). 
The next question is, can such a domain's performance also be \emph{verified} with this same quality,
i.e, will the domain be caught if it lies?

The answer depends, of course, on how many resources the domain's neighbors devote to sampling
and aggregation.
Suppose, for instance, that domain $L$ from Figure~\ref{fig:sla} collects receipts from $X$ and $N$.
Figure~\ref{fig:sampling} gives some concrete numbers: If $X$ samples at $1$\% and 
loses $25$\% of the observed traffic, $L$ can estimate $X$'s delay performance with accuracy $2$msec.
If $N$ samples at the same rate, $L$ can also verify $X$'s performance with the same accuracy.
However, if $N$ samples at $0.1$\%, then $L$ can only verify $X$'s delay performance with accuracy $5$msec.

To summarize, a \Sys domain's choice of sampling and aggregation rate determines,
first, with what quality its own performance can be estimated by its customers and peers;
second, to what extent its receipts can be used to verify the performance of its neighbors.

\section{Discussion and Related Work}
\label{sec:related}

\paragraph*{Partial Deployment}
If domain $X$ in path $\dlp$ has not deployed \Sys, but its neighbors have,
then $X$'s neighbors are free to blame their performance problems on $X$ (since $X$ does not
produce any receipts to refute their claims).  We view this as an incentive for deployment:
a domain has to report on its performance in order to prevent its neighbors
from blaming their problems on it.
Conversely, if $X$ is the only domain in $\dlp$ that \emph{has} deployed \Sys, its performance
reports may not be verified by its neighbors, but they are still verifi\emph{able}.
So, during a congestion incident, $X$ can still position itself as the ``good'' ISP that provides
troubleshooting information to its customers---it is not its fault that the other ISPs on the path 
are not up to the task.
$X$ can even use this as an incentive to encourage multi-network customers to connect all their
networks through $X$---since that way they avoid domains that do not provide troubleshooting
information.

\paragraph*{Related Work}
The Packet Obituaries protocol~\cite{Maniatis04} and the fault-localization protocols
from~\cite{Goldberg07} inform traffic sources where individual packets get lost or corrupted.
AudIt provides source domains with similar \emph{per-TCP-flow} information~\cite{Argyraki07}.
\Sys is similar to these protocols in that it relies on in-path elements collecting 
and exporting traffic statistics; it also borrows the concept of report consistency from AudIt.
\Sys's novel elements are delay-sampling and tunable reporting; based on these techniques,
it avoids the overheads necessary for collecting 
and propagating per-packet or per-flow state, while maintaining the verifiability property.

In Trajectory Sampling, routers within an ISP sample packets using a hash function and record their digests, 
with the purpose of inferring the internal paths (sequences of routers) followed by packets~\cite{Duffield01}.
The Lossy Difference Aggregator enables two monitoring points to measure the loss and average
delay between them by maintaining packet counts and average timestamps for packet aggregates~\cite{Kompella09}.
We use ideas from both protocols (hash-based sampling, per-aggregate counts),
but, as explained in Section~\ref{sec:new}, none of them could provide the computability and verifiability
properties necessary in our context.

The ``Secure Sampling'' technique from~\cite{Goldberg08} is useful when two entities, 
say Alice and Bob, want to measure the 
delay of the path between them by considering only a sample of the packets they exchange.
To prevent intermediate nodes from treating the samples preferentially, Alice and Bob agree on which packets to sample
in such a way that the intermediate nodes cannot guess which are the samples.
This technique is clearly not applicable to our problem:
we are not looking to hide the samples from the intermediate nodes, we are looking to force the intermediate nodes
to sample honestly---in our context, the entities that perform the sampling (the domains) are precisely the ones that 
may bias the samples.

The ``Secure Sketch'' technique from~\cite{Goldberg08} enables Alice and Bob to detect when the packets they exchange 
are lost, delayed, or modified beyond a certain level. To this end, both Alice and Bob maintain a sketch (in some
sense, a summary) of all the packets they have exchanged; at the end, Alice sends her sketch to Bob, who compares
the sketches and detects whether any of the above problems occurred.
This technique is related to \Sys in the same way with the Lossy Difference Aggregator:
we could combine it with the strawman to build a mechanism that determines whether each domain
modified packets beyond a certain level; however, it would not enable the estimation of delay quantiles.

Finally, \Sys can be viewed as a ``performance accountability mechanism,'' which holds domains accountable for their performance.
An economic analysis has showed that such a performance accountability mechanism
would foster ISP competition and innovation~\cite{Laskowski06}.

\iffalse
\Sys shares a similar philosophy with PeerReview, where the nodes of a 
distributed system (any system that can be modeled as a collection of deterministic state machines) 
audit each other for Byzantine behavior~\cite{Haeberlen07}.
Our solution differs in functionality (we are interested in loss and delay performance, 
not Byzantine behavior) and mechanism---in PeerReview, auditing nodes identify faulty peers 
by replaying secure logs of exchanged messages through a reference implementation.

We share similar goals with probing tools that infer ISP performance from end-to-end measurements---from the 
widely used traceroute program to network tomography~\cite{Bu02, Nguyen07c} and the more recent Netdiff~\cite{Mahajan08}
and NVLens~\cite{Zhang08}.
\Sys differs from these techniques in that it produces measurements from actual traffic (not probes)
and, most importantly, in that it targets verifiability---it enables ISPs to produce verifiable 
evidence of their performance.
It also differs in philosophy: instead of treating ISPs as black boxes,
it creates a cooperative framework, where peering networks help verify each other's performance.
\fi

\section{Conclusions}
\label{sec:conclusions}

We have presented \Sys, a system by which network domains can estimate and verify each
other's loss and delay performance. 
\Sys relies on domains producing and exchanging receipts for the traffic they receive and deliver.
A domain can estimate a neighbor's performance by processing the receipts produced by the neighbor;
it can verify that the neighbor's receipts are honest by comparing them to the receipts produced
by other domains for the same traffic.
If a domain lies about its performance, that leads to receipt inconsistencies and exposes the
liar to its neighbors.
\Sys comes at the cost of deploying (modest) new functionality at domain boundaries.
The processing, memory, and bandwidth overhead incurred by a deploying domain is configurable
and independently determined by the domain.

\newpage
{
\footnotesize
\bibliographystyle{abbrv}
\bibliography{bibliography}
}

\end{document}